\documentclass{article}

\usepackage[english]{babel}
\usepackage[utf8]{inputenc}
\usepackage{t1enc}
\usepackage{graphicx}
\usepackage{url}
\usepackage{multirow}
\usepackage{xcolor}
\usepackage{hyperref}
\usepackage{pgfplots}
\usepackage[margin=2.5cm]{geometry}
\usepackage{nameref}
\usepackage{amsmath}

\definecolor{actionable}{HTML}{2F6A28}
\definecolor{non-actionable}{HTML}{C00000}

\title{A Large-Scale Collection Of (Non-)Actionable Static Code Analysis Reports}
\author{
    Dávid Kószó$^1$ \and 
    Tamás Aladics$^{\ast, 1,2}$
    \and 
    Rudolf Ferenc$^1$
    \and 
    Péter Hegedűs$^{1,2}$ 
}

\date{%
    $^\ast$corresponding author, e-mail: \url{aladics@inf.u-szeged.hu}\\[2ex]
    $^1$University of Szeged, Szeged, Hungary\\%
    $^2$FrontEndART Ltd., Szeged, Hungary\\[2ex]%
    \today
}

\begin{document}

\sloppy

\maketitle

\begin{abstract}
Static Code Analysis (SCA) tools, while invaluable for identifying potential coding problems, functional bugs, or vulnerabilities, often generate an overwhelming number of warnings, many of which are non-actionable. 
This overload of alerts leads to ``alert fatigue'', a phenomenon where developers become desensitized to warnings, potentially overlooking critical issues and ultimately hindering productivity and code quality. 
Analyzing these warnings and training machine learning models to identify and filter them requires substantial datasets, which are currently scarce, particularly for Java. 
This scarcity impedes efforts to improve the accuracy and usability of SCA tools and mitigate the effects of alert fatigue. 
In this paper, we address this gap by introducing a novel methodology for collecting and categorizing SCA warnings, effectively distinguishing actionable from non-actionable ones. 
We further leverage this methodology to generate a large-scale dataset of over 1 million entries of Java source code warnings, named NASCAR: (Non-)Actionable Static Code Analysis Reports. 
To facilitate follow-up research in this domain, we make both the dataset and the tools used to generate it publicly available. 


\end{abstract}

\section*{Background and Summary} \label{sect:background_and_summary}

Static code analysis (SCA) tools have become essential in modern software development. 
These tools can contribute significantly to improve software quality and security for the following reasons. 
Firstly, they can catch issues early in the development lifecycle~\cite{avatar, mining_fix_patterns}, and thus, software quality resources (\emph{e.g.}, human costs, \emph{etc.}) can be allocated more effectively. 
Furthermore, since SCA tools can detect quality and security issues in a target program without a process of dynamic execution, they provide an automated way to proactively identify potential bugs and security vulnerabilities~\cite{sa_value}.
To achieve that, SCA tools define the most common bug patterns collected and reviewed by experts in the field, and check the existence of those patterns in target programs. 
According to recent research, the two most popular Java SCA tools \cite{fp_summary} are FindBugs (\url{https://findbugs.sourceforge.net/}, accessed: 2025-02-03), along with its successor, SpotBugs (\url{https://spotbugs.github.io/}, accessed: 2025-02-03) and PMD (\url{https://pmd.github.io/}, accessed: 2025-02-03).

An SCA tool works in the following way. 
Its input is a computer program (either in the form of source code or in the form of bytecode, \emph{etc.}), and its output is a report containing a (possibly empty) list of warnings, \emph{i.e.}, potential flaws of the program. 
Recall that it produces the report without actually running the program, this is why we call it static code analysis. 
Based on the opinion of developers of the computer program, if a warning signals an actual bug that must be repaired, then we call that warning \emph{actionable} (in this paper abbreviated by: A), otherwise we call it \emph{non-actionable} (abbreviated by: NA). Note that some works use the term \emph{false positive} interchangeably with \emph{non-actionable} \cite{d2a, KHARKAR, new_world_dataset}, while others use it in a stricter sense to describe a factual error in the SCA tool's analysis~\cite{fp_summary}. To avoid this ambiguity and to better reflect the real life environment of software development, we define a \emph{non-actionable warning} based on developer action: it is any warning that is systematically not addressed, irrespective of the underlying reason.

Unfortunately, the effectiveness of an SCA tool is often hindered by a high rate of non-actionable warnings. 
For instance, Heckman et al. found that the number of non-actionable warnings is between 35\% and 91\% of the total number of warnings~\cite{HECKMAN2011363}. 
Moreover, manually reviewing and triaging these warnings can be time-consuming, often pulling developers away from their actual coding tasks. 
As a result of this, developers may spend significant time investigating these warnings, which can lead to frustration and reduced adoption of SCA tools. 
This phenomenon is known as ``alert fatigue'', where developers become desensitized to warnings, potentially overlooking critical issues and ultimately hindering productivity and code quality~\cite{KHARKAR,REYNOLDS}.

Machine learning (ML) offers a promising solution to this challenge. 
ML models can be trained to intelligently filter or prioritize warnings, predicting the likelihood that a given warning represents a genuine issue that requires attention~\cite{Yang2020LearningTR}.
This enables developers to focus their efforts on the most critical warnings, thereby improving the efficiency and effectiveness of SCA tools.

The success of ML-based approaches, however, relies heavily on the availability of large and diverse training datasets \cite{fp_summary,schelter_large_scale}.
Unfortunately, in the context of Java SCA, there is a scarcity of publicly available data suitable for this purpose. 
 One of the first such datasets was created with the purpose of benchmarking by Heckman et al.~\cite{Heckman_2008}, called FAULTBENCH. 
 It contains $357$ alerts from $6$ open source projects. 
 Another popular dataset used as a benchmark is generated from the Open Web Application Security Project (OWASP) project (\url{https://owasp.org/www-project-benchmark/}, accessed: 2025-02-03), of which, at the time of writing, two versions are publicly available. 
 Version~1.1 contains $21,041$ test cases, and version~1.2 consists of $2,740$ test cases.
 There are also synthetic datasets of vulnerabilities that can be used for warning ranking or filtering, such as Juliet~\cite{juliet} or SARD (\url{https://samate.nist.gov/SRD/index.php}, accessed: 2025-02-03). 
  Even though these datasets are good quality and manually curated, they are not large enough to train ML models or assess the performance of SCA tools in a real-world setting. 
  Indeed, Chakraborty et al. argued that artificial vulnerabilities are too simple and are not suitable for evaluating the performance of ML models in a real-world setting~\cite{deepl_are_we_there_yet}.

Other works use internal datasets tailored to their specific research questions, which are also often small in size and may not be suitable for reuse. 
For example, Koc et al. collected a dataset from $14$ Java programs, evaluated by FindSecBugs, resulting in $400$ vulnerability warnings~\cite{kocetal1}. 
Ngo et al. created a dataset of $6,620$ warnings in $10$ open source C/C++ projects~\cite{ngo_etal}. 
Lee et al. collected $56,036$ closed source code alarms, which is reduced to $9,871$ after deduplication~\cite{Lee2019ClassifyingFP}. As it can be observed, such internal datasets are limited in size and usually not publicly available.

In this paper, we bridge this data gap as follows.
We introduce a novel approach for collecting and categorizing warnings from SCA tools, effectively distinguishing actionable warnings from non-actionable ones. 
Furthermore, we leverage this approach to generate a substantial new dataset using software repository mining, which is essential for training and evaluating ML models aimed at enhancing the precision and utility of Java SCA.

The ``New Real-World Dataset''~\cite{new_world_dataset} is one of our prior works that has directly impacted this research. 
In that study, we identified the limitations of existing datasets for training ML models to filter non-actionable warnings from SCA tools. 
There we observed that many previous works relied on synthetic benchmarks or small, manually curated datasets (\emph{e.g.} OWASP, \cite{juliet}, \emph{etc.}), which might not accurately reflect the complexities and nuances of real-world codebases. 
To address this issue, we proposed a novel approach for generating a large-scale dataset of real-world SCA warnings. 
The method involved mining the commit histories of open-source Java projects on GitHub, looking for instances where developers had explicitly fixed or ignored warnings by using the \textit{//NOSONAR} directive. 
By leveraging these actions as indicators of actionable and non-actionable warnings, respectively, we were able to construct a dataset that was larger than any previously available public datasets for Java language. 
The dataset contains $224,484$ warning instances from $160$ different SonarQube (\url{https://www.sonarsource.com/products/sonarqube/}, accessed: 2025-02-03) bug patterns across $9,958$ Java projects. 
The data spanned from $2010$, the time when SonarQube entries first appeared in code histories, to the time of the study. 
On the contrary, in this research we analyze fewer Java projects, but collect more actionable and non-actionable warnings, and, in addition, our data is more recent, spanning from $2022$. 
Moreover, we use different SCA tools for identifying warnings, and consider more bug patterns. 
Further details can be found in the \nameref{sect:methodology} section.

The ``Differential Analysis for AI-based Vulnerability Detection'' (D2A)~\cite{d2a} approach presents a contrasting methodology for dataset creation. 
The authors of D2A directly tackle the challenge of non-actionable warnings in SCA tool reports by leveraging the concept of differential analysis on code commits. 
The core idea is to analyze changes between consecutive versions of a project, specifically focusing on bug-fixing commits. 
By running SCA tools on both the before-commit and after-commit versions, D2A identifies issues that disappear in the latter, assuming these to be actionable warnings. 
Issues that persist across versions are then flagged as potential non-actionable ones.

Observe that this approach offers three advantages. 
Firstly, it directly addresses the problem of non-actionable warnings, a major pain point for developers using SCA tools. 
Secondly, it leverages real-world code changes and bug fixes, making the generated dataset more reflective of actual development practices. 
Thirdly, it provides rich contextual information, including bug traces and the specific code changes that resolved the issues.
The D2A dataset was generated by analyzing version pairs from six open-source C/C++ projects. 
The authors employed a commit message analyzer to pre-filter commits likely to be bug fixes, followed by SCA using the Infer tool (\url{https://fbinfer.com/}, accessed: 2025-02-03) on the selected version pairs. 
In their case, Infer reported more than $349$ million issues. 
However, their resulting dataset contained only $1.3$ million unique examples after deduplication. 
For our analysis of open-source Java projects, we employ two SCA tools, in contrast to the single tool used by D2A. We modified their configurations to filter out noise from warnings related to code formatting and style conventions, therefore focusing the analysis on more impactful code quality defects and software security (\emph{cf.} \nameref{sect:methodology} section).
Furthermore, we also obtain a dataset with over a million unique examples, but from fewer issues reported by the SCA tools (\emph{cf.} \nameref{sect:valid} section).

The study presented by Wang et al.~\cite{goldenset} also employs a similar strategy for establishing ground truth labels for warnings. 
The authors state that they utilize a ``commonly-used method'' where the disappearance of a warning in subsequent revisions indicates it was actionable, while the persistence of a warning suggests it was non-actionable. 
This approach, like the one used in D2A, leverages the implicit feedback from developers' actions (or inaction) on warnings to infer their true nature. 
The convergence of these methodologies further strengthens the validity of using code revision history as a reliable indicator of warning actionability.

In our approach, we sought to combine the strengths of both the ``New Real-World Dataset'' and the D2A methodologies. 
We drew inspiration from D2A's innovative use of differential analysis to identify actionable and non-actionable warnings, adapting this concept to the Java language domain. 
Simultaneously, we adopted the ``New Real-World Dataset's'' focus on actionable warning detection and its utilization of real-world Java projects from GitHub as our data source.

\section*{Methods} \label{sect:methodology}

\paragraph{Data collection.}

\begin{figure}[t]
\centering
\includegraphics[scale=0.1]{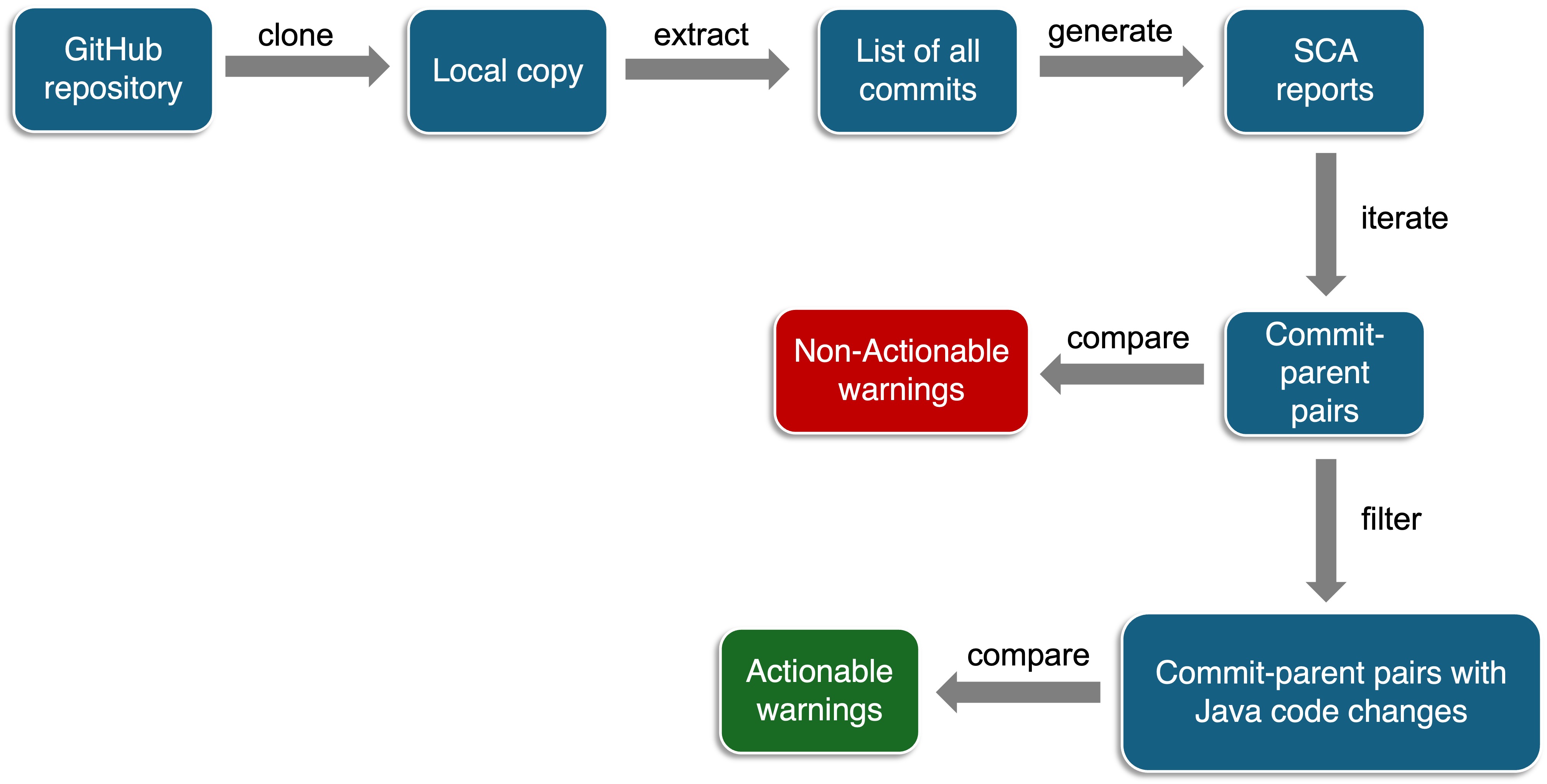}
\caption{\label{fig:flowchart}Flowchart of our data collection process} 
\end{figure}

To explain our data collection process, we give its high-level flowchart (\emph{cf.} Figure~\ref{fig:flowchart}).
In this flowchart, each data block is shown using a rectangle with rounded corners, and each process is depicted by a grey arrow. 

Assume that we are given a GitHub repository URL. 
Firstly, we clone that repository to create its local copy. 
For our analysis, only the main branch of the repository is of interest. 
In the next step, we extract the list of all commits.
Of course, here we have the option to filter out some commits by date using the usual arguments ``since'' or ``until''. 
If neither of the two aforementioned parameters is supplied, then we consider all commits of the main branch. 
In our case, we have supplied the argument ``since'' with the value ``2022-01-01''. 
Next, for each commit in the list, we generate an SCA report. 
Further details on the selected SCA tools can be found later in this section.

Then we consider commit-parent pairs. To improve clarity throughout this section, we will use a consistent notation for each pair. Let $C_c$ denote the current commit and $C_p$ denote its parent commit and their corresponding SCA reports are denoted by $R_c$ and $R_p$, respectively. If a commit $C_c$ does not contain any Java code changes relative to $C_p$, its report $R_c$ will not introduce new actionable warnings, and we do not process that pair further.

When forming ($C_c$, $C_p$) pairs on the main branch, merge commits can easily occur. A merge commit $C_c$ has at least two parents, whereas a regular commit has only one. In our process, the first parent of a merge commit is always on the main branch, while other parents are from different branches. Therefore, if $C_c$ is a merge commit, we define its parent $C_p$ to be its first parent exclusively for our analysis.

Then, to find actionable warnings (\emph{cf.} the green rectangle), we compare the reports $R_c$ and $R_p$. A warning from $R_p$ is marked as actionable if both of the following conditions are met:
\begin{enumerate}
    \item The context of the warning from $R_p$ is affected by a Java code change (an insertion or deletion) in $C_c$.
    \item The warning's message does not appear in the parts of $R_c$ that correspond to the Java code changes in $C_c$.
\end{enumerate}

To find non-actionable warnings (\emph{cf.} the red rectangle), we again compare $R_p$ and $R_c$, this time looking for similarities. A warning from $R_p$ is marked as non-actionable if it also occurs in $R_c$. This persistence can happen for two reasons: either its context was not affected by any Java code change in $C_c$, or its context was affected, but the problem it indicates was not resolved. If a warning is flagged as both actionable and non-actionable, we resolve the conflict by keeping only the actionable classification.
Since, by nature of the SCA tools, the number of non-actionable warnings can be much higher than the number of actionable warnings (\emph{cf.} \nameref{sect:background_and_summary} section), we have decided to reduce the number of non-actionable warnings while reducing the number of exact duplicates: if the same non-actionable warning occurs more than once, then we drop the duplicates, and keep only its last occurrence, \emph{i.e.}, the warning appears in the SCA report of the latest commit. 

An alternative approach to our methodology involves significantly reducing the search space of commits by using VulCurator \cite{vulcurator22}, a tool for identifying vulnerability-fixing commits, which is a mature alternative to the approach used in D2A \cite{d2a}. We integrated this capability into our toolchain, where VulCurator would rank all commits, allowing us to focus our analysis on a TOP n subset most likely to contain security fixes. Our initial experiments in a test environment confirmed the utility of this feature as it successfully filtered out a large number of irrelevant commits.

However, we did not use this feature in the final data collection for this paper due to a critical performance trade-off. The analysis of every commit with VulCurator proved to be too resource-intensive, making it impractical for the large scale of our study, given our goal of collecting a broad and comprehensive dataset in a reasonable time. Nevertheless, the feature remains part of our tool's design, offering an option for future work or for smaller scale studies where depth of analysis is prioritized over breadth and computational time.

Regarding platform-specific details, for data collection we utilized a computer with a $16$-core CPU, $128$ GB RAM, $500$ GB storage, and Ubuntu 20.04.6 LTS.

\paragraph{Selected SCA tools.} \label{subsect:sca}

A crucial step in generating our dataset was the selection of appropriate SCA tools for producing warning reports. 
We began by considering four tools that are commonly utilized in related work~\cite{fp_summary} for analyzing Java source code: SpotBugs, SonarQube, PMD, and Checkstyle (\url{https://checkstyle.sourceforge.io/}, accessed: 2025-02-03).

We conducted preliminary test runs with each aforementioned tool on a sample of Java projects to evaluate the nature and relevance of the warnings generated. 
Following this assessment, we decided to exclude Checkstyle from further consideration because its primary focus is on enforcing code conventions and style-related guidelines, and thus, we continued with SpotBugs, SonarQube, and PMD. 
To utilize SpotBugs and SonarQube, the source code must be successfully compilable, which is a very strict requirement in the case of real-life Java projects. 
On the other hand, PMD does not have this requirement, hence, we decided to work with PMD in the first place.

\begin{figure}
    \centering
    \includegraphics{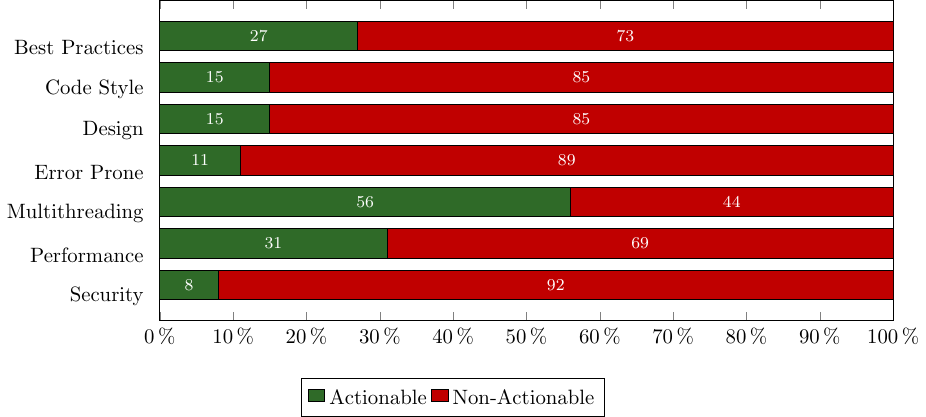}
    \caption{\label{fig:pmd-stat} Distribution of the collected actionable and non-actionable PMD warnings in each bug pattern category except Documentation}
\end{figure}

PMD defines the following eight main bug pattern categories: \href{https://docs.pmd-code.org/latest/pmd_rules_java_bestpractices.html}{Best Practices}, \href{https://docs.pmd-code.org/latest/pmd_rules_java_codestyle.html}{Code Style}, \href{https://docs.pmd-code.org/latest/pmd_rules_java_design.html}{Design}, \href{https://docs.pmd-code.org/latest/pmd_rules_java_documentation.html}{Documentation}, \href{https://docs.pmd-code.org/latest/pmd_rules_java_errorprone.html}{Error Prone}, \href{https://docs.pmd-code.org/latest/pmd_rules_java_multithreading.html}{Multithreading}, \href{https://docs.pmd-code.org/latest/pmd_rules_java_performance.html}{Performance}, and \href{https://docs.pmd-code.org/latest/pmd_rules_java_security.html}{Security}. 
Since rules of the category Documentation are not in the center of our interest, we decided to ignore them.
Hence, in case of PMD we considered $283$ bug patterns in total. 
Figure~\ref{fig:pmd-stat} shows the distribution of the collected actionable and non-actionable PMD warnings in each bug pattern category except Documentation. 
PMD allows us to define our own rules for Java SCA as well, but to keep our results reproducible, we use only built-in rules in our method.

Since the warnings identified by PMD are generally different than the warnings can be identified by SpotBugs and SonarQube, we decided to include one of the two. 
Due to our conducted preliminary test runs, SpotBugs ran a slightly faster than SonarQube. 
Hence, we selected SpotBugs as a second tool.

\begin{figure}
    \centering
    \includegraphics{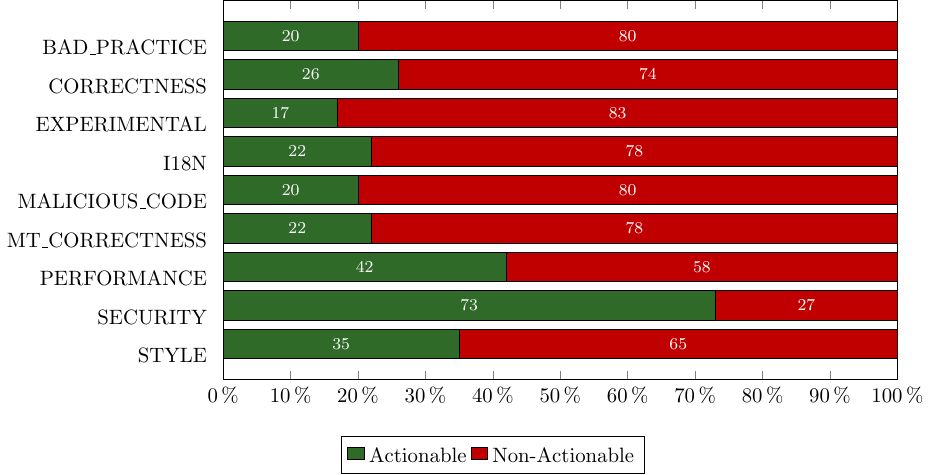}
    \caption{\label{fig:spotbugs-stat} Distribution of the collected actionable and non-actionable SpotBugs warnings in each bug pattern category except NOISE}
\end{figure}

SpotBugs define ten main bug pattern categories, which are as follows: 
\href{https://spotbugs.readthedocs.io/en/latest/bugDescriptions.html#bad-practice-bad-practice}{Bad practice (BAD\_PRACTICE)}, 
\href{https://spotbugs.readthedocs.io/en/latest/bugDescriptions.html#correctness-correctness}{Correctness (CORRECTNESS)}, 
\href{https://spotbugs.readthedocs.io/en/latest/bugDescriptions.html#experimental-experimental}{Experimental (EXPERIMENTAL)}, 
\href{https://spotbugs.readthedocs.io/en/latest/bugDescriptions.html#internationalization-i18n}{Internationalization (I18N)}, 
\href{https://spotbugs.readthedocs.io/en/latest/bugDescriptions.html#malicious-code-vulnerability-malicious-code}{Malicious code vulnerability  (MALICIOUS\_CODE)}, 
\href{https://spotbugs.readthedocs.io/en/latest/bugDescriptions.html#multithreaded-correctness-mt-correctness}{Multithreaded correctness  (MT\_CORRECTNESS)}, 
\href{https://spotbugs.readthedocs.io/en/latest/bugDescriptions.html#bogus-random-noise-noise}{Bogus random noise (NOISE)}, 
\href{https://spotbugs.readthedocs.io/en/latest/bugDescriptions.html#performance-performance}{Performance (PERFORMANCE)}, 
\href{https://spotbugs.readthedocs.io/en/latest/bugDescriptions.html#security-security}{Security (SECURITY)}, and 
\href{https://spotbugs.readthedocs.io/en/latest/bugDescriptions.html#dodgy-code-style}{Dodgy code (STYLE)}. 
Since, we are interested in each of them, none of them was excluded, and thus, in case of SpotBugs we consider $490$ bug patterns in total. 
Figure~\ref{fig:spotbugs-stat} shows the distribution of the collected actionable and non-actionable SpotBugs warnings in each bug pattern category except NOISE. 
We note that no warning is associated with the category NOISE, which is intended to be useful as a control in data collection experiments, not in finding actual bugs in software. 

Referring to the work of ~\cite{fp_summary} and considering our results, we observed that both PMD and SpotBugs were mature and, in addition to that, they are also widely adopted tools in the Java development community.
Their extensive use and established track records ensured that they were likely to generate a diverse and representative set of warnings, making them well-suited for our purposes.

Table~\ref{table:rule-coverage-stats} shows the distribution of bug pattern types (i.e., rules) from PMD and SpotBugs. These types are categorized based on the nature of the individual warning instances they generated within our dataset. A bug pattern type falls into the \emph{both actionable and non-actionable} category if it produced at least one warning classified as actionable and at least one classified as non-actionable in different contexts. This highlights a key finding of our work: the actionability of a warning often depends on the specific code context, not just its type. 

On the other hand, the \emph{neither actionable nor non-actionable} category includes bug pattern types for which no warnings were found and classified in our dataset, either because the rule never triggered on our corpus or because its warnings did not meet our labeling criteria. The data shows that for PMD, 81\% of the observed bug pattern types generated both actionable and non-actionable warnings, emphasizing the strong need for contextual data. The lower percentage for SpotBugs (16\%) is due to a smaller sample of its warnings in our dataset.

\begin{table}[t]
  \small
  \centering
  \caption{\label{table:rule-coverage-stats} 
  Distribution of bug pattern categories defined by PMD and SpotBugs with respect to the warning classes}
  \begin{tabular}{lrrrr}
    \multirow{2}{*}{\textbf{Warning Classes}} & \multicolumn{2}{c}{\textbf{PMD}} & \multicolumn{2}{c}{\textbf{SpotBugs}}\\\cline{2-5}
    & \multicolumn{1}{c}{\textbf{Quantity}} & \multicolumn{1}{c}{\textbf{Percentage}} & \multicolumn{1}{c}{\textbf{Quantity}} & \multicolumn{1}{c}{\textbf{Percentage}}\\
    \hline
    \hline
    Both A and NA &  $231$ & $81.63\%$ & $79$ & $16.12\%$\\
    Only A & $0$ & $0.00\%$  & $21$ & $4.29\%$\\
    Only NA & $20$ & $7.07\%$  & $119$ & $24.29\%$\\
    Neither A nor NA & $32$ & $11.30\%$  & $271$ & $55.31\%$\\
    \hline
    \textbf{Sum} & $\mathbf{283}$ & $\mathbf{100.00\%}$ & $\mathbf{490}$ & $\textbf{100.00\%}$\\
    \hline
  \end{tabular}
\end{table}

\paragraph{Projects.}\label{sect:projects}

For our purpose, we have carefully selected $102$ Java projects available on GitHub based on the number of stars and activity to ensure data quality. 
More precisely, we have selected a project if it has had at least $200$ stars in GitHub and it has been updated in the last two years, given that we performed the data collection in November 2024.
If, in addition, the latest version of a project, \emph{i.e.}, the version after the last commit, can be compiled successfully using Maven or Gradle, then it is considered a significant advantage as it can also be analyzed with SpotBugs in addition to PMD. 

Table~\ref{table:proj-stats} shows twenty of the selected 102 projects. 
The first ten projects are the biggest ones (in terms of the number of commits), for which we have found at least one warning. 
Recall that, in case of a project, it is difficult to run SpotBugs, because each commit must be compilable, which does not always hold true. 
Moreover, even if that holds true, it is very time-consuming to perform each compilation. 
Since PMD does not have such high requirements, in case of repositories with more than 1000 commits, we only run PMD, \emph{i.e.}, we have found zero SpotBugs warnings for such projects. 
Then we give the ten biggest projects in terms of the number of identified SpotBugs warnings. 
Observe that those ten projects can be considered rather small in terms of the number of commits.

\begin{table}[t]
    \small
    \centering
	\caption{\label{table:proj-stats}Project statistics of our new data collection process}
	\begin{tabular}{lrrrrrrr}
		\multirow{2}{*}{\textbf{Project Name}}  & \multirow{2}{*}{\textbf{Commits}} & \multirow{2}{*}{\textbf{Stars}} & \multirow{2}{*}{\textbf{Forks}} & \multicolumn{2}{c}{\textbf{PMD}} & \multicolumn{2}{c}{\textbf{SpotBugs}} \\\cline{5-8}
		&&&& \multicolumn{1}{c}{\textbf{A}} & \multicolumn{1}{c}{\textbf{NA}} & \multicolumn{1}{c}{\textbf{A}} & \multicolumn{1}{c}{\textbf{NA}}\\
		\hline
		\hline
		\href{https://github.com/spring-projects/spring-framework}{spring-framework} & $31,892$ & $57,072$ & $38,286$ & $2,489$ & $0$ & $0$ & $0$\\
	    \href{https://github.com/orientechnologies/orientdb}{orientdb} & $25,944$ & $4,762$ & $871$ & $17,136$ & $576$ & $0$ & $0$ \\
	    \href{https://github.com/OpenAPITools/openapi-generator}{openapi-generator} & $20,835$ & $22,386$ & $6,665$ & $14,069$ & $619$ & $0$ & $0$\\
	    \href{https://github.com/Anuken/Mindustry}{Mindustry} & $18,296$ & $23,048$ & $3,011$ & $656$ & $32,007$ & $0$ & $0$ \\
	    \href{https://github.com/apache/maven}{maven} & $15,171$ & $4,441$ & $2,693$ & $3,525$ & $18,933$ & $0$ & $0$\\
	    \href{https://github.com/apache/kafka}{kafka} & $14,669$ & $29,190$ & $14,080$ & $5,694$ & $609$ & $0$ & $0$\\
	    \href{https://github.com/runelite/runelite}{runelite} & $14,203$ & $4,867$ & $5,266$ & $969$ & $112,441$ & $0$ & $0$\\
	    \href{https://github.com/flowable/flowable-engine}{flowable-engine} & $13,607$ & $8,117$ & $2,641$ & $17,855$ & $69,912$ & $0$ & $0$\\
	   \href{https://github.com/jOOQ/jOOQ}{jOOQ} & $13,593$ & $6,231$ & $1,212$ & $4,723$ & $749$ & $0$ & $0$\\
	   \href{https://github.com/knowm/XChange}{XChange} & $13,313$ & $3,899$ & $1,949$ & $19,651$ & $22,334$ & $0$ & $0$\\
	   \hline
	   \href{https://github.com/alibaba/nacos}{nacos} &	$5,155$ & $30,610$  &	$12,901$ &	$10,869$ & 	$24,340$	 & $473$ &	$97$\\
	   \href{https://github.com/hneemann/Digital}{Digital} & $4,548$ & $4,555$&  $456$ & $50$  & $15,793$  	& $2$	& $442$	\\
	   \href{https://github.com/jfinal/jfinal}{jfinal}	& $1,655$ &  $3,241$ &	$1,314$ & 	$422$ & 	$6,529$  & 	$13$ & 	$239$ \\		
	   \href{https://github.com/zhkl0228/unidbg}{unidbg} &	$1,662$ & $4,020$	& $978$  & 	$494$ & $16,165$ &	$29$ & $221$\\	
       \href{https://github.com/sofastack/sofa-jraft}{sofa-jraft} &	$380$ & $3,633$ & 	$1,154$ 		& $297$  & $8,712$ 	& $2$ & $244$ \\
       \href{https://github.com/FasterXML/jackson-databind}{jackson-databind} & $8,082$ & $3,536$	& $1,390$ & $2,259$ & $24,940$	&	$7$  & $210$	\\	
	   \href{https://github.com/LibrePDF/OpenPDF}{OpenPDF} &	$1,612$ & $3,668$ &	$603$	&  $4,281$ & $17,901$ & $38$ & $174$	\\
	   \href{https://github.com/Atmosphere/atmosphere}{atmosphere}	& $6,110$ &  $3,700$	& $751$	& $471$ &  $5,521$ & $36$ & $162$\\
	   \href{https://github.com/bytedeco/javacpp}{javacpp} &	$947$	 & $4,524$	& $588$ &	$45$ & $6,878$	&	$5$ & $167$	\\		
       \href{https://github.com/networknt/light-4j}{light-4j} & $2,502$ & $3,619$ & $633$ & $4,459$ & $9,852$ & $86$ & $78$\\	   
       \vdots & \vdots & \vdots & \vdots & \vdots & \vdots & \vdots & \vdots\\
        \hline
        \textbf{Sum} & $\mathbf{528,912}$ & $\mathbf{1,079,001}$ & $\mathbf{313,798}$ & $\mathbf{195,766}$ & $\mathbf{1,027,376}$ & $\mathbf{1,174}$ & $\mathbf{3,447}$\\
        \hline
	\end{tabular}
\end{table}

\section*{Data Records} \label{sect:data-records}
The datasets with the corresponding source files can be reached here~\cite{nascar}: \\ \url{https://doi.org/10.5281/zenodo.17079912}

The replication package contains two main directories, \texttt{dataset} and \texttt{tools}.

\begin{figure}
    \centering
    \includegraphics{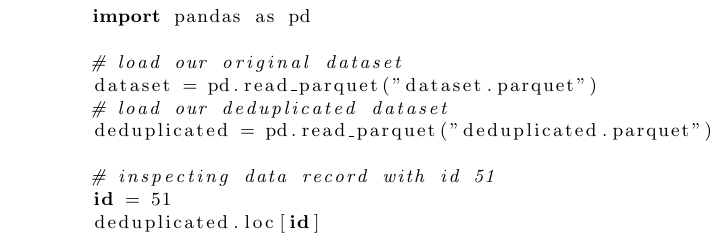}
    \caption{\label{fig:usage} Code sample for processing our dataset}
\end{figure}

The \texttt{dataset} directory consists of two parts, (1) actionability data stored in parquet format(\url{https://parquet.apache.org}, accessed: 2025-02-03) and (2) the corresponding Java source code files. The actionability data is stored in two parquet files, \texttt{dataset.parquet} and \texttt{deduplicated.parquet}, both of which can be processed \emph{e.g.} with pandas (\emph{cf.} Figure~\ref{fig:usage}). \texttt{dataset.parquet} is the main dataset, containing all of the $1,227,763$ data records ($196,940$ ($16\%$) actionable and $1,030,823$ ($84\%$) non-actionable warnings (\emph{cf.} Table~\ref{table:proj-stats}), while \texttt{deduplicated.parquet} contains the deduplicated version, a strict subset of the main dataset, containing $1,083,073$ records. For details regarding the deduplication, please refer to the \nameref{sect:valid} section. For each warning we also include a complete copy of the source file that we store in a directory structure compressed and stored in a ZIP archive, \texttt{files.zip}. If a source file belongs to more than one warning, then we keep just one copy of that source file.

The \texttt{tools} directory contains the \texttt{miner} and \texttt{deduplication} directories, which contain the source code of our data collection and deduplication tools, respectively.

\paragraph{Structure.}
Table~\ref{table:dataset-structure} describes the structure of our dataset. 
Each field name in the first column of the table corresponds to a column name in our dataset. 
In case of some columns (\emph{e.g.}, \texttt{label}) further restrictions may apply, which we mention in the fourth column of the table.

\begin{table}[t]
    \small
    \centering
    \caption{\label{table:dataset-structure} Structure of our dataset}
    \vspace*{1em} 
    \begin{tabular}{llp{5cm}p{4cm}}
    \textbf{Field Name} & \textbf{Field Type} & \textbf{Field Description} & \textbf{Restrictions}\\
    \hline
    \hline
    \texttt{tool} & string & the name of the SCA tool, which created the warning & only \texttt{PMD} or \texttt{SpotBugs}\\
    \hline
    \texttt{warning\_type} & string & the name of the violated rule of the SCA tool & built-in Java rules of PMD and built-in rules of SpotBugs\\
    \hline
    \texttt{warning\_msg} & string & the detailed explanation of the warning\\
    \hline
    \texttt{parent\_sha} & string & the unique hash of the parent commit in which the warning was reported \\
    \hline
    \texttt{parent\_date} & date & the date of the parent commit converted to UTC and formatted to ISO 8601 \\
    \hline
    \texttt{commit\_sha} & string & the unique hash of the commit following the parent commit \\
    \hline
    \texttt{commit\_date} & date & the date of the commit converted to UTC and formatted to ISO 8601 \\
    \hline
    \texttt{repo} & string & the url of the project's GitHub repository\\
    \hline
    \texttt{filename} & string & the name of the source file, in which the warning occurred\\
    \hline
    \texttt{positions} & stringified JSON & the start and the end lines of the narrowest context within the source file, in which the warning occurred; the start and the end columns sometimes are also given\\
    \hline
    \texttt{filepath} & string & the path to the source file within our dataset\\
    \hline
    \texttt{label} & int & the label of the data record; \texttt{0} for non-actionable and \texttt{1} for actionable & only \texttt{0} or \texttt{1}\\
    \hline
    \end{tabular}
\end{table}

\section*{Technical Validation} \label{sect:valid}

\subsection*{Deduplication}
Since we have analyzed hundreds of commits of several projects, it is natural that the same warning has occurred multiple times either in the same project or in a different project.
This statement is particularly true for non-actionable warnings, but, in some cases, it holds true for actionable warnings as well. 
Note that these warnings may only differ in their contexts. 
To reduce the number of such duplicates, we apply a state-of-the-art method: deduplication \cite{d2a}. 
For this, we perform the min-wise independent permutations locality sensitive hashing scheme (for short: MinHash and LSH) \cite{bro97}, which is a well-known technique for quickly estimating how similar two finite sets are.

MinHash and LSH use the Jaccard similarity coefficient to perform their computations. 
The Jaccard similarity coefficient expresses the similarity between two finite sets, and is defined as the size of the intersection divided by the size of the union of the two sets. 
Its value is a real number between $0$ and $1$, where $0$ indicates that the two sets are disjoint, and $1$ indicates that they are equal. 
In particular, the two sets are more similar, \emph{i.e.}, have relatively more common members if their Jaccard similarity coefficient is closer to $1$. 
However, in our case, since we look for duplicates, the lower the value of the Jaccard similarity coefficient is, the more duplicates can be found. 
Moreover, to determine similarities, MinHash and LSH also need permutation functions, which are applied to the sets to obtain random permutations of set elements. 
Evidently, the higher the number of permutation functions is, the more thorough the calculation is but the more time is needed for the calculation. 

In our case, for each warning, we consider three lines before (respectively, after) the start (respectively, end) line of the warning if applicable. 
We set the Jaccard similarity coefficient to $0.95$. 
Moreover, we set the number of permutation functions to $128$. 
After the deduplication process, $144,690$ duplicated data records have been identified, which yields a deduplicated dataset with $1,083,073$ records, among which there are $145,997$ ($13,48\%$) actionable and $937,076$ ($86,52\%$) non-actionable warnings
(the original dataset consists of $1,227,763$ records, \emph{cf.} \nameref{sect:methodology} section). 
The deduplication process is reproducible, because the random permutation functions are generated deterministically for MinHash by default.

\subsection*{Manual Validation}
Furthermore, to ensure the validity of the data, we perform a manual validation of our deduplicated dataset. 
Evidently, manual validation of each warning in our deduplicated dataset cannot be carried out due to the huge amount of data, so we have validated only some selected samples.
To determine the minimum sample size required for a statistically significant manual validation, we employed the standard formula for a proportion from a finite population. In more detail, the calculation is based on Cochran's formula with a finite population correction, as follows:
$$n = \frac{n_0}{1 + \frac{n_0 - 1}{N}} \quad \text{where} \quad n_0 = \frac{z^2 \cdot \hat{p}(1-\hat{p})}{\epsilon^2}$$
The parameters for this calculation were chosen as follows:
\begin{itemize}
    \item \textbf{Population Size ($N$):} We used the full size of our deduplicated dataset, $N = 1,083,073$. Since this number is large, the finite population correction factor will have a minimal effect on the sample size.
    \item \textbf{Confidence Level:} We selected a 90\% confidence level, a common choice considering the constraints of manual analysis, which corresponds to a z-score ($z$) of 1.645.
    \item \textbf{Margin of Error ($\epsilon$):} We set the margin of error to 10\% ($\epsilon = 0.10$).
    \item \textbf{Population Proportion ($\hat{p}$):} As we have no knowledge of the real-world distribution of actionable warnings, we used a population proportion of 50\% ($\hat{p} = 0.5$). This is the most conservative estimate, as it maximizes the variance and yields the largest required sample size.
\end{itemize}

Applying these parameters to the formula results in a required sample size of 69 warnings.  By randomly selecting and validating this number of warnings, we can generalize our findings to the entire dataset with the specified statistical constraints. Finally, two of the authors have manually validated those $69$ samples independently (the list of these entries can be found in the replication package~\cite{nascar}). 

Our manual validation strategy has been straightforward. In case of each sample warning, we wished not to overrule the decision of the maintainers of the projects what is actionable and what is not as they can judge those questions better than us. We have just simply checked the Java code changes of the commits in the following way. If a sample warning has been labelled as actionable by our data miner, then we have compared the relevant commit  ($C_c$) with its parent ($C_p$) and looked for the existence of a particular Java code change, which can resolve the problem. If we have found such code change, then that actionable warning has been marked as passed, and as failed otherwise. On the other hand, if a sample warning has been labelled as non-actionable by our data miner, then we have compared $C_c$ with $C_p$ and looked for the non-existence of any particular Java code changes, which can handle the signaled problem. If we have not found any such code change, then that non-actionable warning has been marked as passed, and failed otherwise. Both authors have found that all the $69$ sample warnings have been marked as passed.

\subsection*{Threats to Validity}
A potential threat to construct validity lies in our measurement of non-actionable warnings. As discussed in the \nameref{sect:background_and_summary} section, we define a non-actionable warning as any warning not addressed by developers, irrespective of the underlying reason. This definition opens up the possibility that a warning's persistence may not come from an explicit developer decision, but from their unawareness of the warning's existence, particularly if they do not use PMD or SpotBugs in their regular workflow. This could mean that our dataset of non-actionable warnings is a result of developer ignorance rather than conscious disregard. However, we argue this threat is mitigated by several factors in our study design and goals.

 Our research is explicitly motivated by the goal of understanding and combating ``alert fatigue'' in a real-world context. From this perspective, a warning that is systematically ignored, even due to a lack of tool integration or developer awareness, is functionally non-actionable and contributes to the overall problem we aim to study. Furthermore, our methodology labels a warning as non-actionable only if it is \emph{systematically} unaddressed across multiple sequential commit pairs (pre- and post-commit states, as described in the \nameref{sect:methodology} section). This also means that the analysis potentially spans the work of multiple developers, reducing the likelihood that a warning's persistence is due to a single developer's inattention or unawareness.

This is complemented by our focus on high-quality, actively maintained projects (as mentioned in the \nameref{sect:projects} section), which increases the probability that developers know of common software engineering best practices, including the use of static analysis. 

Finally, automatically determining which specific SCA tools a project integrates is a significant technical challenge, as there is no standard convention for this in version control systems. Our approach is therefore consistent with the methodology of related works in this area, as they also do not perform specific filtering based on SCA tool usage on a project basis~\cite{fp_summary, d2a, new_world_dataset}.

\begin{table}[t]
    \small
    \centering
    \caption{\label{table:sample-selection} Statistics of sample selection}
    \vspace*{1em} 
    \begin{tabular}{llrrrr}
    \multirow{2}{*}{\textbf{SCA}} & \multirow{2}{*}{\textbf{Warning Type}} & \multicolumn{2}{c}{\textbf{Deduplicated Dataset}} & \multicolumn{2}{c}{\textbf{Representative Sample}}\\\cline{3-6}
    & & \multicolumn{1}{c}{\textbf{Quantity}} & \multicolumn{1}{c}{\textbf{Percentage}} & \multicolumn{1}{c}{\textbf{Quantity}} & \multicolumn{1}{c}{\textbf{Percentage}}\\
    \hline
    \hline
    \multirow{2}{*}{PMD} & Actionable & $145,019$ & $13.39\%$ & $10$ & $14.49\%$ \\
    & Non-Actionable & $934,030$ & $86.24\%$ & $57$ & $82.61\%$\\
     \multirow{2}{*}{SpotBugs} & Actionable & $978$ & $0.09\%$ & $0$ &  $0.00\%$\\
      & Non-Actionable & $3,046$  & $0.28\%$ & $2$ & $2.90\%$\\
      \hline
      \multicolumn{2}{c}{\textbf{Sum}} & $\mathbf{1,083,073}$ & $\mathbf{100.00\%}$ & $\mathbf{69}$ & $\mathbf{100.00\%}$
    \end{tabular}
\end{table}

\section*{Usage Notes}
A sequential, platform-independent procedure for dataset reproduction is provided below. For more in-depth details, please refer to the README file in the respective directories~\cite{nascar}.
\begin{enumerate}
    \item Begin by installing all tools and libraries listed in the \nameref{sect:code-availability} section, as they are required for successful execution of the workflow.
    \item Retrieve the published artifacts from Zenodo \cite{nascar}.
    \item Launch a terminal and change to the root directory of the miner tool.
    \item Ensure that execute permissions are set for the Python scripts to be run (\emph{e.g.} \texttt{chmod a+x} \emph{etc.}). 
    \item Set up the environment with the appropriate Python libraries (\emph{cf.} \texttt{requirements.txt}).
    \item The analysis is launched. This can be done for a single repository using the \texttt{miner.py} script or for a list of repositories (in a .txt file where each line contains a target Git repository) using the \texttt{feeder.py} script. The analysis can be optionally filtered by a specific commit date range, using the \texttt{---since} and \texttt{---until} arguments.
    \item After the mining is complete, the collected source files are checked using the \texttt{file\_checker.py} script.
    \item Subsequently, the \texttt{create\_dataset.py} script is run to generate the final dataset in a Parquet database file and to organize the collected source code into a structured directory.
    \item Finally, different statistics about the projects and warnings are computed by running the \texttt{proj\_stats.py}, \texttt{pmd\_stats.py}, and \texttt{spotbugs\_stats.py} scripts.
    \item Optionally, to obtain the deduplicated version of the dataset, execute the \texttt{dedup\_nascar.py} script located in the \texttt{deduplication} directory.
\end{enumerate}

\section*{Data Availability} \label{sect:data-availability}
The datasets with the corresponding source files are available here: \\ \url{https://doi.org/10.5281/zenodo.17079912}

\section*{Code Availability} \label{sect:code-availability}
To produce the NASCAR dataset, we have written several Python scripts. 
These scripts can be found in the directory \texttt{miner} of our repository~\cite{nascar}. 
Moreover, to obtain \texttt{deduplicated.parquet} from  \texttt{dataset.parquet}, we also have written small Python scripts, which are stored in the directory \texttt{deduplication}. 

The dataset was generated by executing our generator scripts using the following software: \texttt{Ubuntu 20.04.6 LTS}, \texttt{Python 3.10.12}, \texttt{OpenJDK 17.0.15}, \texttt{Apache Maven 3.9.6}, \texttt{Gradle 8.9}, \texttt{PMD 7.1.0}, and \texttt{SpotBugs 4.8.3}.

\bibliographystyle{naturemag}

\section*{Author Contributions }
Dávid Kószó: Conceptualization, Writing - Original Draft, Data Collection, Technical Validation, Visualization.
Tamás Aladics: Conceptualization, Technical Validation, Writing - Draft Preparation. Rudolf Ferenc: Project Coordination, Writing - Review \& Editing. 
Péter Hegedűs: Conceptualization, Project Coordination, Writing - Review \& Editing.

All authors read, edited, and approved the final manuscript.

\section*{Competing Interests}
The authors declare that they have no competing interests.

\section*{Acknowledgements}
This work was supported in part by the European Union project RRF-2.3.1-21-2022-00004 within the framework of the Artificial Intelligence National Laboratory; and in part by the Project no TKP2021-NVA-09 implemented with the support provided by the Ministry of Culture and Innovation of Hungary from the National Research, Development and Innovation Fund, financed under the TKP2021-NVA funding scheme. 
The work has also received support from the European Union Horizon Program under the grant number 101120393 (Sec4AI4Sec).

\end{document}